\documentclass[conference]{IEEEtran}
\IEEEoverridecommandlockouts
\usepackage{cite}
\usepackage{amsmath,amssymb,amsfonts}
\usepackage[ruled,linesnumbered,lined]{algorithm2e}
\usepackage{graphicx}
\usepackage{textcomp}
\usepackage{xcolor}
\usepackage{bm}
\usepackage{stfloats}
\usepackage{url}
\usepackage{verbatim}
\usepackage[caption=false,font=normalsize,labelfont=sf,textfont=sf]{subfig}
\usepackage{algorithmic}

\usepackage{tcolorbox}
\newtcbox{\mybox}{colback=white,colframe=black,boxrule=1pt}
\begin{document}

\title{Transmitter Side Beyond-Diagonal RIS for mmWave Integrated Sensing and Communications\\
\thanks{This work has been supported in part by the National Nature Science Foundation of China under Grant 62201347; and in part by Shanghai Sailing Program under Grant 22YF1428400. (\textit{Corresponding author: Yijie Mao})} 
\vspace{-5pt}
}

\author{\IEEEauthorblockN{Kexin Chen, Yijie Mao}
\IEEEauthorblockA{School of Information Science and Technology, ShanghaiTech University, Shanghai 201210, China \\
Email: \{chenkx2023, maoyj\}@shanghaitech.edu.cn}
}
\maketitle

\begin{abstract}
% BD-RIS aided transmitter architecture
This work initiates the study of a beyond-diagonal reconfigurable intelligent surface (BD-RIS)-aided transmitter architecture for integrated sensing and communication (ISAC) in the millimeter-wave (mmWave) frequency band.
% the purpose
Deploying BD-RIS at the transmitter side not only alleviates the need for extensive fully digital radio frequency (RF) chains but also enhances both communication and sensing performance. These benefits are facilitated by the additional design flexibility introduced by the fully-connected scattering matrix of BD-RIS.
% the algorithm
To achieve the aforementioned benefits, in this work, we propose an efficient two-stage algorithm to design the digital beamforming of the transmitter and the scattering matrix of the BD-RIS with the aim of jointly maximizing the sum rate for multiple communication users and minimizing the largest eigenvalue of the Cram{\'e}r-Rao bound (CRB) matrix for multiple sensing targets.
% the simulation
Numerical results show that the transmitter-side BD-RIS-aided mmWave ISAC outperforms the conventional diagonal-RIS-aided ones in both communication and sensing performance.
\end{abstract}


\section{Introduction}
% integrated sensing and communication (ISAC) 
Integrated sensing and communication (ISAC) has emerged as a critical enabler for next-generation wireless networks. This attributes to its potential for sharing the spectrum, hardware architectures, and signal processing modules between communication and sensing functionalities. %\cite{lf22_tutorial}. 
% millimeter-wave (mmWave)
Meanwhile, the incorporation of millimeter-wave (mmWave) technique opens the door to high data rates for communications %\cite{Bua18_mmWave} 
and high-resolution capabilities for target sensing \cite{lf22_tutorial}. %\cite{cyh21_mmWave}. 
Therefore, mmWave holds great promise for ISAC systems.
However, owing to the severe path loss caused by the short wavelength characteristics of mmWave, a large number of transmit antennas along with extensive use of fully digital radio frequency (RF) chains are normally required at the transmitter to achieve high beamforming gain, resulting in huge power consumption \cite{jamali21_activeantenna}. This  calls for low-cost solutions for mmWave ISAC systems.

One promising solution for mmWave ISAC is utilizing the reconfigurable intelligent surfaces (RIS) to assist the transmission. RIS, composed of numerous passive elements, shows its capability to reconfigure the wireless propagation environment for communication and sensing functionalities in the mmWave band \cite{zzy22_mmWaveISAC,lrw23_mmWaveISAC,ly24_mmWaveISAC}. %Specifically, \cite{zzy22_mmWaveISAC} demonstrated the benefits of RIS in enhancing the transmission rate of mmWave ISAC systems. And \cite{lrw23_mmWaveISAC} minimized the trace of the CRB matrix to enhance estimation accuracy in RIS-aided mmWave ISAC networks, considering hybrid analog-digital beamforming. Additionally, catering to practical mmWave systems, \cite{ly24_mmWaveISAC} explored a simultaneous training and sensing protocol assisted by RIS. 
Moreover, RIS adjusts the phase of the incident signals at an ultra-low power cost. It eliminates the need for extensive RF processing at the transmitter\cite{jamali21_activeantenna}. 
% the disadvantage
However, most of the existing studies on RIS-aided mmWave ISAC \cite{zzy22_mmWaveISAC,lrw23_mmWaveISAC,ly24_mmWaveISAC} primarily concentrate on the receiver-side RIS with more emphasis on the former benefit of RIS in propagation reconfiguration. 
%There is a lack of study on transmitter-side RIS for mmWave ISAC systems. 

In addition, a revolutionary RIS architecture named beyond-diagonal RIS (BD-RIS) has been recently proposed \cite{ssp22_BDRIS, lhy22_BDRIS}. %BD-RIS can be categorized into fully-, group- and single-connected architectures, each offering different trade-offs between the hardware complexity and system performance \cite{lhy22_BDRIS}. 
Different from conventional diagonal-RIS (D-RIS) where each element operates independently, in fully-connected BD-RIS \cite{lhy22_BDRIS}, all elements are connected to each other. %With the aid of BD-RIS, both the phase and magnitude of incident signals can be adjusted. 
Recent studies \cite{WBW23_BDRISISAC, esmaeil24_BDRISISAC} have shown the superior performance gain of the receiver-side BD-RIS over conventional D-RIS in terms of both communication and sensing performance. However, the transmitter-side BD-RIS for ISAC system and its application in the mmWave frequency band has not been investigated yet.

Inspired by the transmitter-side RIS-aided communication network introduced in \cite{jamali21_activeantenna,dwn23_ITScommunication,Anup2024BDRIS}, in this work, we initiate the study of a transmitter-side BD-RIS-aided mmWave ISAC system. Specifically, we design the digital beamforming of the transmitter and the scattering matrix of the BD-RIS to jointly maximize the communication sum rate and minimize the largest eigenvalue of the sensing Cram{\'e}r-Rao bound (CRB). An efficient two-stage optimization method is proposed, where the scattering matrix of the BD-RIS is obtained via the symmetric unitary projection, and the digital beamforming is optimized subsequently via the successive convex approximation (SCA) method. Numerical results demonstrate that BD-RIS-aided mmWave ISAC achieves a better communication and sensing trade-off compared to the conventional D-RIS-aided ones.
%%ITS
%%%%%%%%%%%%%%%%%%%%%%%%%%%%%%%%%%%%%%%%%%%%%%%%%%%%%%%%
%% the system model figure
\begin{figure}
    \centering
    \includegraphics[width=1\linewidth]{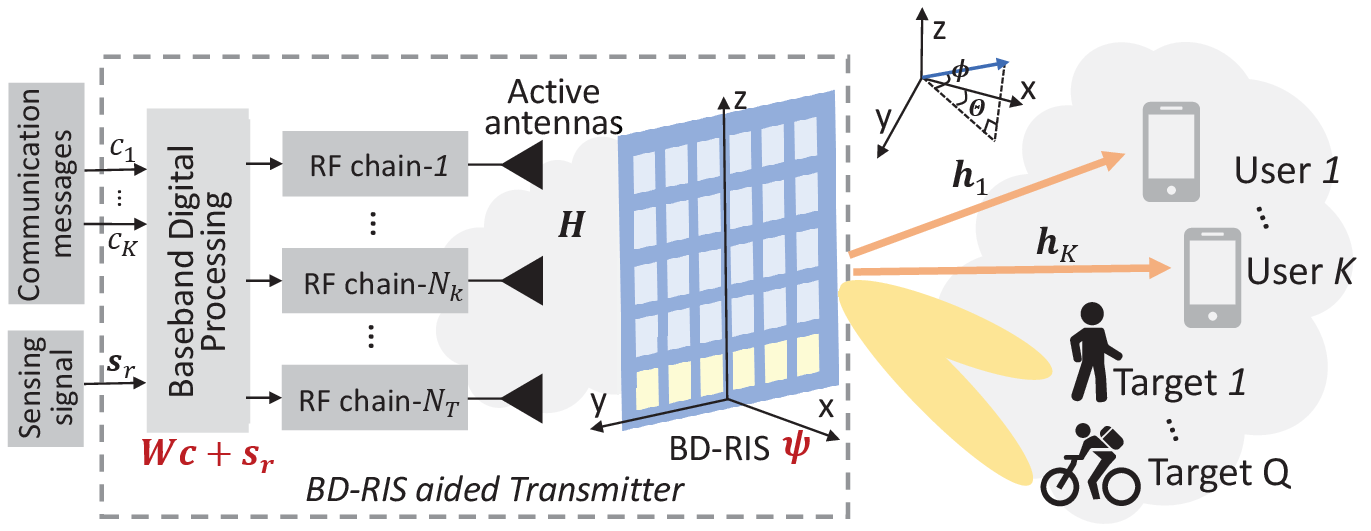}
    \vspace{-0.7cm}
    \caption{The system model of BD-RIS-aided transmitter architecture for ISAC.}
    \label{model}
    \vspace{-0.5cm}
\end{figure}
%
%%%%%%%%%%%%%%%%%%%%%%%%%%%%%%%%%%%%%%%%%%%%%%%%%%%%%%%%%%%%%%%%%%%%%%%%%%%%%%%%%%%%%%%%%%%%%%%%%%%%%%%%%%%%%%%%
\section{System Model} 
As depicted in Fig. \ref{model}, we consider a mmWave ISAC system with a BD-RIS-aided transmitter simultaneously serving  $K$ single-antenna communication users and $Q$ sensing targets indexed by $\mathcal{K}=\left \{ 1,\dots, K \right \}$ and $\mathcal{Q}=\left \{ 1,\dots, Q \right \}$, respectively. 
At the transmitter, the $N_T$ active feed antennas are connected to $N_T$ dedicated RF chains, and the dual-functional signal is transmitted directly to a BD-RIS with $N_I$ elements. 
% the channel model
The fully-connected BD-RIS considered in this work operates in the transmissive mode\footnote{We solely focus on the transmissive mode of BD-RIS. This choice is based on its demonstrated advantage of experiencing less feed blockage when positioned a few wavelengths away from the active antennas, as opposed to the reflective mode \cite{jamali21_activeantenna}.}, where all $N_I$ elements are connected to each other and allow incident signals to penetrate them only \cite{lhy22_BDRIS}. %\cite{jamali21_activeantenna, bai20_ITS}.
Moreover, BD-RIS is assumed to be deployed at a short distance, typically a few wavelengths away from the active antennas. The modeling of the channel matrix $\mathbf{H} \in \mathbb{C}^{N_I \times N_T}$ between the active antennas and the BD-RIS follows the uniform separate illumination (SI) strategy \cite{jamali21_activeantenna,dwn23_ITScommunication}. And the scattering matrix of the fully-connected BD-RIS is a full matrix denoted by $\bm{\Psi} \in \mathbb{C}^{N_I \times N_I}$. It satisfies the following symmetric and unitary constraint \cite{ssp22_BDRIS, lhy22_BDRIS}:
\begin{equation}\label{psi_S}
    \bm{\Psi}\in\mathcal{S}, \mathcal{S}=\{\bm{\Psi}|\bm{\Psi}=\bm{\Psi}^T, \bm{\Psi}^H\bm{\Psi}=\mathbf{I}_{N_I}\}.
\end{equation}

A dedicated low-cost sensor with $N_S$ elements is placed at the BD-RIS to receive the echo signals from the three-hop link \cite{wzl23_STARS}, i.e., the active antennas at the transmitter$\to$BD-RIS$\to$targets$\to$sensor. This reduces the number of hops and path loss as compared with the conventional receiver-side RIS-aided systems, where the sensing functionality is carried out at the transmitter far away from the RIS. 
Moreover, we assume that the side of the sensor facing the active antennas is physically blocked in order to achieve unidirectional reception of echo signals \cite{wzl23_STARS}.
We consider $M$ transmission blocks indexed by $\mathcal{M}=\{1,\dots ,M\}$ in one coherent processing interval (CPI). The data streams for communication users at time index $m$ are given as $\mathbf{c}[m]=\left[c_{1}[m],\dots,c_{K}[m]\right]^{T}\in \mathbb{C}^{K\times 1}$. $ \mathbf{W}=[\mathbf{w}_{1},\dots, \mathbf{w}_{K}]\in \mathbb{C}^{N_{T}\times K}$ denotes the corresponding digital precoding matrix. It remains consistent during one CPI. 
A dedicated sensing signal $\mathbf{s}_r[m] \in \mathbb{C}^{N_{T}\times 1}$ is considered in this work. Its covariance matrix is $\mathbf{R}_S=\mathbb{E}\left[\mathbf{s}_r[m]\mathbf{s}_r^H[m]\right]$. The transmit signal at time index $m$ is given by
\begin{equation}
	\mathbf{x}[m]=\mathbf{W}\mathbf{c}[m]+\mathbf{s}_r[m]=\sum_{k\in \mathcal{K}}\mathbf{w}_{k}c_{k}[m]+ \mathbf{s}_r[m],
\end{equation}
where the communication and sensing streams satisfy $\mathbb{E}[\mathbf{c}[m]\mathbf{c}^{H}[m]]=\mathbf{I}_{K}$, and $\mathbb{E}[\mathbf{c}[m]\mathbf{s}^{H}_r[m]]=\mathbf{0}_{K\times N_T}$, implying that the entries of the communication and sensing signals are independent from each other. The covariance matrix of the transmit signal is thereby given as $\mathbf{R}_{X}=\frac{1}{M}\sum_{m\in \mathcal{M}}\mathbf{x}[m]\mathbf{x}[m]^{H}=\mathbf{W}\mathbf{W}^{H}+\mathbf{R}_S$.
%%%%%%%%%%%%%%%%%%%%%%%%%%%%%%%%%%%%%%%%%%%%%%%%%%%%%%%%    
%% communication signal

\textit{Communication users:}
The signal received at the communication user $k$ at time index $m$ is modeled as
\begin{equation} \label{y_com}
	\begin{aligned}
		y_{k}[m]&=\mathbf{h}_{k}^{H}\mathbf{\Psi}\mathbf{H}\mathbf{w}_{k}c_{k}[m]+\sum_{i\in \mathcal{K}, i\neq k}\mathbf{h}_{k}^{H}\mathbf{\Psi}\mathbf{H} \mathbf{w}_{i}c_{i}[m]\\
                &+ \mathbf{h}_{k}^{H}\mathbf{\Psi}\mathbf{H}\mathbf{s}_r[m]+z_{k}[m], \forall k\in \mathcal{K},
	\end{aligned}
\end{equation}
where $\mathbf{h}_{k}\in \mathbb{C}^{N_{I}\times 1}$ denotes the channel vector between the BD-RIS and user $k$, it is assumed to be perfectly known at the transmitter and user $k$. $z_{k}[m]$ is the additive white Gaussian noise (AWGN) with a distribution of $\mathcal{CN}(0,\sigma _{c}^{2})$. 

The signal to interference plus noise ratio (SINR) for decoding the desired signal of user $k$ is given by
\begin{equation}
    \gamma_k=\frac{\left|\mathbf{h}_{k}^H\mathbf{\Psi}\mathbf{H}\mathbf{w}_{k}\right|^2}{\sum_{i\in \mathcal{K}, i\neq k}\left|\mathbf{h}_{k}^H\mathbf{\Psi}\mathbf{H} \mathbf{w}_{i}\right|^2+\mathbf{h}_{k}^H\mathbf{\Psi}\mathbf{H}\mathbf{R}_S\mathbf{H}^{H}\mathbf{\Psi}^{H}\mathbf{h}_{k}+\sigma_c^2}.
\end{equation}
Therefore, the sum rate for communication users is calculated as $\mathrm{R_{sum}}=\sum_{k=1}^{K}\log_2(1+\gamma_k)$. This metric is selected to evaluate the communication performance in this work.
%%%%%%%%%%%%%%%%%%%%%%%%%%%%%%%%%%%%%%%%%%%%%%%%%%%%%%%%
%% sensing signal

\textit{Sensing targets:}
The dual-functional signal $\mathbf{x}[m]$ is also exploited for target sensing. The received echo signal at the sensor from $Q$ targets at time index $m$ is given as
\begin{equation}\label{Y_radar}
	\mathbf{y}_{r}[m]=\mathbf{B}\mathbf{U}\mathbf{A}^{T}\mathbf{\Psi}\mathbf{H}\mathbf{x}[m]+\mathbf{z}_{r}[m],
\end{equation}
where
\begin{equation}\label{Y_r_brief}
	\begin{aligned}
		&\bm{\alpha}=[\alpha _{1},\dots,\alpha _{Q}]^{T}, \mathbf{B}=[\mathbf{b}(\theta _1, \phi_1),\dots,\mathbf{b}(\theta _{Q}, \phi_Q)],\\
		&\mathbf{U}=\mathrm{diag}(\boldsymbol{\alpha}), \mathbf{A}=[\mathbf{a}(\theta _1, \phi_1),\dots,\mathbf{a}(\theta _{Q}, \phi_Q)], \\
	\end{aligned}
\end{equation}
$\alpha_{q}$ in $\bm{\alpha}$ is a complex amplitude determined by the complex reflection coefficient and the path-loss of the $q$th target. $\theta_q$ and $\phi_q$ respectively denote the $q$th target's azimuth and elevation angle of departure (DoA) to the BD-RIS. $\mathbf{z}_{r}[m] \in \mathbb{C}^{N_{S}\times 1}$ is the AWGN following $\mathcal{CN}(\mathbf{0}_{N_{S}\times 1},\sigma _{r}^{2}\mathbf{I}_{N_{S}})$. Let $[\bm{r}_x, \bm{r}_y, \bm{r}_z] \in \mathbb{R}^{N_I \times 3}$ and $[\hat{\bm{r}}_x, \hat{\bm{r}}_y, \hat{\bm{r}}_z] \in \mathbb{R}^{N_S \times 3}$ respectively denote the Cartesian coordinates of the BD-RIS and the sensor. With the BD-RIS deployed in $yz$ plane and the sensor along the $y$ axis, i.e., $\bm{r}_x=\bm{0}$, $\hat{\bm{r}}_x=\hat{\bm{r}}_z=\bm{0}$, the steering vector can be calculated as
$\mathbf{a}(\theta _q, \phi_q)=e^{-j\frac{2\pi }{\lambda}(\mathbf{r}_y\sin(\theta_q)\cos(\phi_q)+\mathbf{r}_z \sin(\phi_q))}$, and $\mathbf{b}(\theta _q, \phi_q)=e^{-j\frac{2\pi }{\lambda} (\hat{\mathbf{r}}_y\sin(\theta_q)\cos(\phi_q))}$, where $\lambda$ denotes the wavelength of the carrier frequency.
%\begin{equation}
%    \begin{aligned}
 %       \mathbf{a}(\theta _q, \phi_q)&=e^{-j\frac{2\pi }{\lambda} %(\mathbf{r}_y\sin(\theta_q)\cos(\phi_q)+\mathbf{r}_z \sin(\phi_q))},\\
%        \mathbf{b}(\theta _q, \phi_q)&=e^{-j\frac{2\pi }{\lambda} %(\hat{\mathbf{r}}_y\sin(\theta_q)\cos(\phi_q))}.
%    \end{aligned}
%\end{equation}

%The angles $\theta_q$ and $\phi_q$ can be obtained via the Capon method \cite{Jardak17_Capon}. 
The CRB matrix is exploited to measure the estimation performance. It is equivalent to the inverse of the fisher information matrix (FIM) $\mathbf{F}$ \cite{lj08_CRB}.
$\mathbf{F} \in \mathbb{R}^{4Q \times 4Q}$ for estimating the parameter set $\mathbf{\xi }_{q}=\left \{ \theta_{q} ,\phi_{q}, \mathrm{Re}(\alpha_q), \mathrm{Im}(\alpha_q) \right \}^{T}$, $\forall q\in\mathcal{Q}$ is
\begin{equation}\label{fisher}
	\mathbf{F}=\begin{bmatrix}
		\mathbf{F}_{\bm{\theta}\bm{\theta}} & \mathbf{F}_{\bm{\theta}\bm{\phi}} & \mathbf{F}_{\bm{\theta}\mathrm{Re}(\bm{\alpha})} & \mathbf{F}_{\bm{\theta}\mathrm{Im}(\bm{\alpha})} \\ 
		\mathbf{F}_{\bm{\theta}\bm{\phi}}^T & \mathbf{F}_{\bm{\phi}\bm{\phi}} & \mathbf{F}_{\bm{\phi}\mathrm{Re}(\bm{\alpha})} & \mathbf{F}_{\bm{\phi}\mathrm{Im}(\bm{\alpha})} \\  
		\mathbf{F}_{\bm{\theta}\mathrm{Re}(\bm{\alpha})}^T & \mathbf{F}_{\bm{\phi}\mathrm{Re}(\bm{\alpha})}^T & \mathbf{F}_{\mathrm{Re}(\bm{\alpha})\mathrm{Re}(\bm{\alpha})} & \mathbf{F}_{\mathrm{Re}(\bm{\alpha})\mathrm{Im}(\bm{\alpha})} \\  
		\mathbf{F}_{\bm{\theta}\mathrm{Im}(\bm{\alpha})}^T & \mathbf{F}_{\bm{\phi}\mathrm{Im}(\bm{\alpha})}^T & \mathbf{F}_{\mathrm{Re}(\bm{\alpha})\mathrm{Im}(\bm{\alpha})}^T & \mathbf{F}_{\mathrm{Im}(\bm{\alpha})\mathrm{Im}(\bm{\alpha})}
	\end{bmatrix},
\end{equation}
where the entries of $\mathbf{F}$ are derived in the Appendix A.
%%%%%%%%%%%%%%%%%%%%%%%%%%%%%%%%%%%%%%%%%%%%%%%%%%%%%%%%%%%%%%%%%%%%%%%%%%%%%%%%%%%%%%%%%%%%%%%%%%%%%%%%%%%%%%%%
\section{Problem Formulation and Optimization Framework}
% RIS coefficient optimization(max the sum of channel gain)
% precoder optimization(max the sum rate and CRB parameter)
In this section, to enhance both the communication and sensing performance of the proposed BD-RIS-aided mmWave ISAC system model,  we design the digital beamforming $\mathbf{W}$ of the transmitter and the scattering matrix $\bm{\Psi}$ of the BD-RIS. Instead of proposing complex joint optimization methods for $\mathbf{W}$ and $\bm{\Psi}$, in this work, we propose a novel two-stage algorithm, where $\bm{\Psi}$ and $\mathbf{W}$ are separately optimized as follows.
%which involves the following two stages:
%\begin{itemize}
%    \item \textit{Stage 1}: In the first stage, we design the coefficient matrix $\bm{\Psi}$ to maximize the sum of the effective channel gains among communication users and sensing targets based on a symmetric unitary projection method \cite{fty24_BDRISlowcomplexity}.
%    \item \textit{Stage 2}: With the optimized $\bm{\Psi}$, we then design the linearly precoding matrix $\mathbf{W}$ to jointly maximize the sum rate for communication users and minimize the largest eigenvalue of the CRB matrix for sensing targets in the second stage. A successive convex approximation (SCA)-based approach is proposed to solve the problem.
%\end{itemize}

\subsection{\textit{Stage 1:} Proposed BD-RIS Scattering Matrix Design} \label{psi section}
%
%Inspired by the approach proposed in \cite{fty24_BDRISlowcomplexity} for a BD-RIS-aided communication-only network, in this subsection, we extend it to a ISAC scenario. To obtain a feasible solution of the coefficient matrix $\bm{\Psi}$, we initially relax the non-convex set $\mathcal{S}$ mentioned in (\ref{psi_S}) onto a convex sphere set $\mathcal{M}=\{\bm{\Psi}|\|\bm{\Psi}\|_F^2\leq N_I\}$, where $\mathcal{S}\in\mathcal{M}$. The relax problem to maximize the sum of the channel gains is formulated as
%\begin{subequations}\label{p1}
%	\begin{align}
% (\mathcal{P}_1)\,\, \mathop{\max}_{\mathbf{\Psi}}\,\,\, &f(\bm{\Psi})=\|\mathbf{G}^H\bm{\Psi}\mathbf{H} \|_F^2 \\
%		s.t.\,\,\,\,\,\,&\bm{\Psi}\in \mathcal{M}, 
%	\end{align}
%\end{subequations}
%where $\mathbf{G}=[\mathbf{h}_1, \dots, \mathbf{h}_K, \mathbf{A}^\ast\mathbf{U}^H\mathbf{B}^H]\in\mathbb{C}^{N_I \times (K+N_S)}$.
Building upon the approach introduced in \cite{fty24_BDRISlowcomplexity} for a BD-RIS-aided communication-only network, this subsection extends its application to an ISAC scenario where the BD-RIS scattering matrix is designed to jointly optimize the sum channel gains of the communication users and sensing targets.
To obtain a feasible solution of the scattering matrix $\bm{\Psi}$, we first relax the non-convex set $\mathcal{S}$ in (\ref{psi_S}) to the set $\mathcal{M}=\{\bm{\Psi}|\bm{\Psi}^H\bm{\Psi}=\mathbf{I}_{N_I}\}$, where $\mathcal{S}\in\mathcal{M}$. The relaxed problem to maximize the sum channel gains is formulated as
\begin{subequations}\label{p1}
	\begin{align}
 (\mathcal{P}_1)\,\, \mathop{\max}_{\mathbf{\Psi}}\,\,&f(\bm{\Psi})=\sum_{k=1}^K\|\mathbf{h}_k^H\bm{\Psi}\mathbf{H} \|^2 +\|\mathbf{B}\mathbf{U}\mathbf{A}^T\bm{\Psi}\mathbf{H} \|_F^2 \label{p1_object}\\
		s.t.\,\,\,\,&\bm{\Psi}\in \mathcal{M}.
	\end{align}
\end{subequations}
The objective function can be further rewritten as $f(\bm{\Psi})=\|\mathbf{G}^H\bm{\Psi}\mathbf{H}\|_F^2$, where $\mathbf{G}=[\mathbf{h}_1, \dots, \mathbf{h}_K, \mathbf{A}^\ast\mathbf{U}^H\mathbf{B}^H]\in\mathbb{C}^{N_I \times (K+N_S)}$. To derive the optimal solution of (\ref{p1}), we first define the singular value decomposition (SVD) of the matrix $\mathbf{G}^H$ and $\mathbf{H}$ as $\mathbf{G}^H=\mathbf{U}_1\mathbf{S}_1\mathbf{V}_1^H$ and $\mathbf{H}=\mathbf{U}_2\mathbf{S}_2\mathbf{V}_2^H$, where $\mathbf{U}_i$, $\mathbf{V}_i$ are unitary matrices while $\mathbf{S}_i$ are diagonal matrices, $i=\{1,2\}$. 

\par
        \textbf{Proposition 1.} \textit{Define $\mathbf{X}=\mathbf{V}^H_1\bm{\Psi}\mathbf{U}_2 \in \mathbb{C}^{N_I \times N_I}$, $\mathbf{\Sigma}_1=\mathbf{S}_1^H\mathbf{S}_1 \in \mathbb{C}^{N_I \times N_I}$, and $\mathbf{\Sigma}_2=\mathbf{S}_2\mathbf{S}_2^H \in \mathbb{C}^{N_I \times N_I}$. Then, we obtain that $\mathrm{Re}\{\mathrm{tr}(\mathbf{\Sigma}_1\mathbf{X}\mathbf{\Sigma}_2\mathbf{X}^H) \}\leq \mathrm{tr}(\mathbf{\Sigma}_1\mathbf{\Sigma}_2)$, with equality holding if and only if $\mathbf{X}=\mathbf{I}_{N_I}$.} \par
        \textit{Proof:} See Appendix B. $\hfill\blacksquare$
    
\par 
Based on Proposition 1, the objective function $f(\bm{\Psi})$ in (\ref{p1_object}) becomes 
\begin{equation}
    \begin{aligned}
        f(\bm{\Psi})&=\|\mathbf{U}_1\mathbf{S}_1\mathbf{V}_1^H\bm{\Psi}\mathbf{U}_2\mathbf{S}_2\mathbf{V}_2^H\|_F^2\\
&=\mathrm{Re}\{\mathrm{tr}(\mathbf{\Sigma}_1\mathbf{X}\mathbf{\Sigma}_2\mathbf{X}^H) \}\\ &\leq \mathrm{tr}(\mathbf{\Sigma}_1\mathbf{\Sigma}_2),
    \end{aligned}
\end{equation}
where the equality holds if and only if $\mathbf{X}=\mathbf{I}_{N_I}$. Therefore, the optimal solution for the relaxed problem (\ref{p1}) is given by
\begin{equation}\label{psi_relaxed}
    \bm{\Psi}^\star=\mathbf{V}_1\mathbf{U}_2^H.
\end{equation}

%\textit{\textbf{Proposition 1} (optimal solution of (\ref{p1}) \cite{Demir22_optimal})\textbf{.}} 
%Through vectorization, we rewrite the objective function of (\ref{p1}) as $f(\bm{\Psi})=\|\mathbf{E}\mathrm{vec}(\bm{\Psi})\|^2$, where $\mathbf{E}=\mathbf{H}^T \otimes \mathbf{G}^H$. Problem (\ref{p1}) can then be optimally solved, of which the optimal solution is
%\begin{equation}\label{psi_relaxed}
%    \mathrm{vec}(\bm{\Psi}^\ast)=\sqrt{N_I}\mathbf{p},
%\end{equation}
%where $\mathbf{p}$ is the normalized dominant eigenvector of $\mathbf{E}^H\mathbf{E}$. Note that $\otimes$ and $\mathrm{vec}(\cdot)$ refer to the Kronecker product and the vectorization operation, respectively. 

%\textit{Proof:} See [\cite{Demir22_optimal}, Appendix A].

Note that the optimal $\bm{\Psi}^\star$ for problem (\ref{p1}) may not belong to the feasible set $\mathcal{S}$ of the BD-RIS. The symmetric unitary projection proposed in \cite{fty24_BDRISlowcomplexity} is then utilized to project a feasible solution on the set $\mathcal{S}$. 
Specifically, let $\tilde{\bm{\Psi}}=\frac{\bm{\Psi}^\star+(\bm{\Psi}^\star)^T}{2}$, and suppose that it has a rank of $g$. Its SVD is $\tilde{\bm{\Psi}}=\mathbf{U}\mathbf{S}\mathbf{V}^H$, where $\mathbf{U}=[\mathbf{U}_{g}\, \mathbf{U}_{N_I-g}]$ and $\mathbf{V}=[\mathbf{V}_{g}\,\mathbf{V}_{N_I-g}]$, respectively. A feasible scattering matrix $\bm{\Psi}$ on the set $\mathcal{S}$ is then calculated as \cite{fty24_BDRISlowcomplexity}
\begin{equation}\label{psi}
    \bm{\Psi}=\hat{\mathbf{U}}\mathbf{V}^H,
\end{equation}
where $\hat{\mathbf{U}}=[\mathbf{U}_{g}, \mathbf{V}^\ast_{N_I-g}]$.
(\ref{psi}) is known as the symmetric unitary projector which projects $\bm{\Psi}^\star$ to $\mathcal{S}$.
%\textit{\textbf{Definition 1} (the symmetric unitary projection \cite{fty24_BDRISlowcomplexity})\textbf{.}} The projection operator $\mathrm{symuni}(\cdot)\to\mathcal{S}$ onto the manifold $\mathcal{S}$ is defined as $\mathrm{symuni}(\mathbf{Z})=\arg \min_{\hat{\mathbf{Q}}\in\mathcal{S}}\|\mathbf{Z}-\hat{\mathbf{Q}}\|_F^2$.
%\begin{equation}
%        \mathrm{symuni}(\mathbf{Z})=\arg \min_{\hat{\mathbf{Q}}\in\mathcal{S}}\|\mathbf{Z}-\hat{\mathbf{Q}}\|_F^2.
%\end{equation}

%Suppose that $\tilde{\mathbf{Z}}=\frac{\mathbf{Z}+\mathbf{Z}^T}{2}$ has the rank of $g$, and $\tilde{\mathbf{Z}}=\mathbf{U}\mathbf{S}\mathbf{V}^H$, where $\mathbf{U}=[\mathbf{U}_{g}\, \mathbf{U}_{N_I-g}]$ and $\mathbf{V}=[\mathbf{V}_{g}\,\mathbf{V}_{N_I-g}]$, respectively. The symmetric unitary projection is then given by
%\begin{equation}
%    \mathrm{symuni}(\mathbf{Z})=\hat{\mathbf{U}}\mathbf{V}^H,
%\end{equation}
%where $\hat{\mathbf{U}}=[\mathbf{U}_{g}, \mathbf{V}^\ast_{N_I-g}]$.

%According to the Definition 1, we derive a feasible scattering matrix $\bm{\Psi}$ on the set $\mathcal{S}$, which is calculated as
%\begin{equation}\label{psi}
%    \bm{\Psi}=\mathrm{symuni}(\bm{\Psi}^\star).
%\end{equation}
\subsection{\textit{Stage 2:} Proposed Transmit Beamforming Design}
With the scattering matrix $\bm{\Psi}$ obtained from (\ref{psi}), we then optimize the digital beamforming matrix $\mathbf{W}$ with the aim of jointly maximizing the sum rate of the communication users and minimizing the largest eigenvalue of the CRB matrix of the sensing targets. 
Note that minimizing the largest eigenvalue of the CRB matrix of the sensing targets is equivalent to maximizing the smallest eigenvalue of $\mathbf{F}$\cite{lj08_CRB}. 
By defining $\mathbf{W}_{u,k}=\mathbf{w}_k\mathbf{w}_k^H$, where $\mathrm{rank}(\mathbf{W}_{u,k})=1, \forall k \in\mathcal{K}$ and introducing $R_k$ as an auxiliary variable to denote the achievable rate of user $k$, $k \in \mathcal{K}$, the optimization problem is formulated as
\begin{subequations}\label{p2}
	\begin{align}
(\mathcal{P}_2)\,\,&\mathop{\max}_{\{\mathbf{W}_{u,k}\}_{k=1}^{K}, r, \mathbf{R}_X, \{R_k\}_{k=1}^K}\,\,\, \sum_{k=1}^K R_k+\mu r \\
		s.t.\,\,\,\,\,\,&\mathbf{F}\succeq  r\mathbf{I}_{4Q}, \label{opt 1}\\
		&\mathrm{tr}(\mathbf{R}_X)\leq P, \label{opt 2}\\
        &\mathbf{R}_X \succeq \sum_{k=1}^K\mathbf{W}_{u,k},\,\mathbf{W}_{u,k}\succeq \mathbf{0}, \forall k \in \mathcal{K}, \label{opt 3}\\
        &\hspace{-2em}\log_2\bigg(1+\frac{\mathbf{v}_k^H\mathbf{W}_{u,k}\mathbf{v}_k}{\mathbf{v}_k^H\mathbf{R}_X\mathbf{v}_k-\mathbf{v}_k^H\mathbf{W}_{u,k}\mathbf{v}_k+\sigma_c^2}\bigg)\geq R_k, \forall k \in \mathcal{K},\label{opt 4}\\
        &\mathrm{rank}(\mathbf{W}_{u,k})=1, \forall k \in \mathcal{K}, \label{opt 5}
	\end{align}
\end{subequations}
where $\mathbf{v}_k^H=\mathbf{h}_k^H\bm{\Psi}\mathbf{H}, \forall k \in \mathcal{K}$ represents the effective channel between the transmitter and user $k$.  $\mu$ is a parameter used to shift the priority between the communication and sensing functionality. The matrix ($\mathbf{F}-r\mathbf{I}_{4Q}$) is guaranteed to be positive semi-definite by constraint (\ref{opt 1}), where $r$ corresponds to the smallest eigenvalue of $\mathbf{F}$. (\ref{opt 2}) is the power constraint of the active antennas, with $P$ denoting the total power budget. Note that the problem is non-convex owing to the rate expressions (\ref{opt 4}) and the rank-one constraints (\ref{opt 5}). 

\begin{algorithm}[tb]
\begin{algorithmic}
    \STATE \hspace{-2em} \textit{\textbf{Stage 1}:} Design the scattering matrix $\bm{\Psi}$. \\
    \STATE Calculate the optimal $\bm{\Psi}^\star$ on the relaxed set $\mathcal{M}$ via (\ref{psi_relaxed}).\\
    \STATE Derive a feasible $\bm{\Psi}$ on the set $\mathcal{S}$ via (\ref{psi}).\\
    \STATE \hspace{-2em} \textit{\textbf{Stage 2}:} Design the transmit beamforming $\mathbf{W}$. \\
    \STATE \textbf{Initialize}: $t\leftarrow0$, $\{\mathbf{W}_{u,k}^{[t]}\}_{k=1}^{K}$, $\bm{\epsilon}^{[t]}$, $\bm{\delta}^{[t]}$, $\bm{\eta}^{[t]}$ \;
    \REPEAT
        \STATE Solve problem (\ref{p3}) with $\{\mathbf{W}_{u,k}^{[t]}\}_{k=1}^{K}$, $\bm{\epsilon}^{[t]}$, $\bm{\delta}^{[t]}$, $\bm{\eta}^{[t]}$, \\and obtain the optimal $\{\mathbf{W}_{u,k}^{\star}\}_{k=1}^{K}$, $\bm{\epsilon}^{\star}$, $\bm{\delta}^{\star}$, $\bm{\eta}^{\star}$ and\\ the optimal objective function value $f_w^{\star}$\;
        \STATE $t\leftarrow t+1$\;
        \STATE Update $\{\mathbf{W}_{u,k}^{[t]}\}_{k=1}^{K}\leftarrow\{\mathbf{W}_{u,k}^{\star}\}_{k=1}^{K}$, $\bm{\epsilon}^{[t]}\leftarrow\bm{\epsilon}^{\star}$, $\bm{\delta}^{[t]}\leftarrow\bm{\delta}^{\star}$, $\bm{\eta}^{[t]}\leftarrow\bm{\eta}^{\star}$, $f_w^{[t]}\leftarrow f_w^{\star}$\;
    \UNTIL {$|f_w^{[t]}-f_w^{[t-1]}|<\varepsilon$}
\end{algorithmic}
\caption{Beamforming optimization with BD-RIS}
\label{algor}
%\vspace{-3mm}
\end{algorithm}
By introducing slack variables $\bm{\epsilon}=[\epsilon_1,\dots, \epsilon_K]$ and $\bm{\delta}=[\delta_1, \dots, \delta_K]$, we first rewrite (\ref{opt 4}) as
\begin{subequations}\label{rate}
    \begin{align}
        &\epsilon_{k}- \delta_{k}\geq R_k \ln2, \forall k \in \mathcal{K},\label{rate_all}\\
        &\mathrm{e}^{\epsilon_k}\leq \mathbf{v}_k^H\mathbf{R}_X\mathbf{v}_k+\sigma_{c}^{2}, \forall k \in \mathcal{K},\label{rate_up}\\
        &\mathrm{e}^{\delta_k}\geq \mathbf{v}_k^H\mathbf{R}_X\mathbf{v}_k-\mathbf{v}_k^H\mathbf{W}_{u,k}\mathbf{v}_k+\sigma_{c}^{2}, \forall k \in \mathcal{K}.\label{rate_down}
    \end{align}    
\end{subequations}
To deal with the non-convexity in (\ref{rate_down}), we leverage the first-order Taylor approximation to approximate the left-hand side (LHS) of (\ref{rate_down}) at the point $\delta_k^{[t]}$. (\ref{rate_down}) is therefore approximated as
\begin{equation}\label{ratedown_tay}
    \mathrm{e}^{\delta_k^{[t]}}(1+\delta_k-\delta_k^{[t]}) \geq  \mathbf{v}_k^H\mathbf{R}_X\mathbf{v}_k-\mathbf{v}_k^H\mathbf{W}_{u,k}\mathbf{v}_k+\sigma_{c}^{2}, \forall k \in \mathcal{K},
\end{equation}
where $t$ denotes the $t$th SCA iteration. 
To achieve more efficient calculation via CVX, the slack variables $\bm{\eta}=[\eta_1, \dots, \eta_K]$ are further introduced to rewrite the non-linear terms $\mathrm{e}^{\epsilon_k}, \forall k \in \mathcal{K}$ in (\ref{rate_up}), which is given by
\begin{subequations}
    \begin{align}
    &\eta_k \leq \mathbf{v}_k^H\mathbf{R}_X\mathbf{v}_k+\sigma_{c}^{2}, \forall k \in \mathcal{K}, \label{rateup_simple}\\
    & \eta_k \ln \eta_k \geq \eta_k \epsilon_k, \forall k \in \mathcal{K} \label{rateup_extra}.
    \end{align}
\end{subequations}
Similarly, $\eta_k \ln \eta_k$ on the LHS of (\ref{rateup_extra}) is approximated at the point $\eta_k^{[t]}, \forall k \in \mathcal{K}$, and then transformed to its equivalent second-order cone (SOC) form as
%\begin{equation}\label{rateup_soc}
%		\left \| \left[\eta_k+\epsilon_k-(1+\mathrm{ln}\eta_{k}^{[t]}), 2\sqrt{\eta_{k}^{[t]}} \right]\right \|_{2}\leq \eta _{k}-\epsilon _{k}+(1+\mathrm{ln}\eta_{k}^{[t]}).
%\end{equation}
\begin{equation}\label{rateup_soc}
\begin{aligned}
    &\left \| \left[\eta_k+\epsilon_k-(1+\mathrm{ln}\eta_{k}^{[t]}), 2\sqrt{\eta_{k}^{[t]}} \right]\right \|_{2}\\
    &\leq \eta _{k}-\epsilon _{k}+(1+\mathrm{ln}\eta_{k}^{[t]}), \forall k \in \mathcal{K}.
\end{aligned}
\end{equation}

With the normalized dominant eigenvector $\mathbf{p}_{k, max}^{[t]}$ of the matrix $\mathbf{W}_{u,k}^{[t]}, \forall k \in \mathcal{K}$, we then move the non-convex rank-one constraints (\ref{opt 5}) to the objective function as %$f_{\mathrm{rank}}=\Gamma \sum_{k\in \mathcal{K}}\left(\mathrm{tr}(\mathbf{W}_{u,k})-(\mathbf{p}_{k,max}^{[t]})^{H}\mathbf{W}_{u,k}\mathbf{p}_{k,max}^{[t]}\right)$, 
\begin{equation}\label{orank}
    f_{\mathrm{rank}}=\Gamma \sum_{k\in \mathcal{K}}\left(\mathrm{tr}(\mathbf{W}_{u,k})-(\mathbf{p}_{k,max}^{[t]})^{H}\mathbf{W}_{u,k}\mathbf{p}_{k,max}^{[t]}\right),
\end{equation}
where $\Gamma$ is a negative constant selected in order to guarantee that $f_{\mathrm{rank}}$ approaches zero closely.

Based on (\ref{rate})--(\ref{orank}), problem (\ref{p2}) is approximated to a sequence of convex subproblems. Specifically, at iteration $t+1$, the following subproblem is tackled with $\mathbf{W}_{u,k}^{[t]}$, $\bm{\epsilon}^{[t]}$, $\bm{\delta}^{[t]}$, $\bm{\eta}^{[t]}$ obtained for the $t$th iteration:
\begin{equation}\label{p3}
	\begin{aligned}
        (\mathcal{P}_3)\,\,&\mathop{\max}_{\{\mathbf{W}_{u,k}\}_{k=1}^K, \mathbf{R}_X, \{R_k\}_{k=1}^K,r, \bm{\epsilon}, \bm{\delta}, \bm{\eta}}\,\,\,\sum_{k=1}^K R_k+\mu r+f_{\mathrm{rank}} \\
		s.t.\,\,\,\,&\textrm{(\ref{opt 1})-(\ref{opt 3}), (\ref{rate_all}), (\ref{ratedown_tay}), (\ref{rateup_simple}), (\ref{rateup_soc}).}
	\end{aligned}
\end{equation}
Problem (\ref{p3}) is convex, and it can be solved by CVX directly.
We summarize the overall optimization process in Algorithm \ref{algor}, where $\varepsilon$ is the convergence tolerance.
\begin{figure*}[t] 
	\vspace{-4mm}
	\begin{minipage}[t]{0.34\linewidth} 
		\centering
		\includegraphics[width=2.2in, height=1.95in]{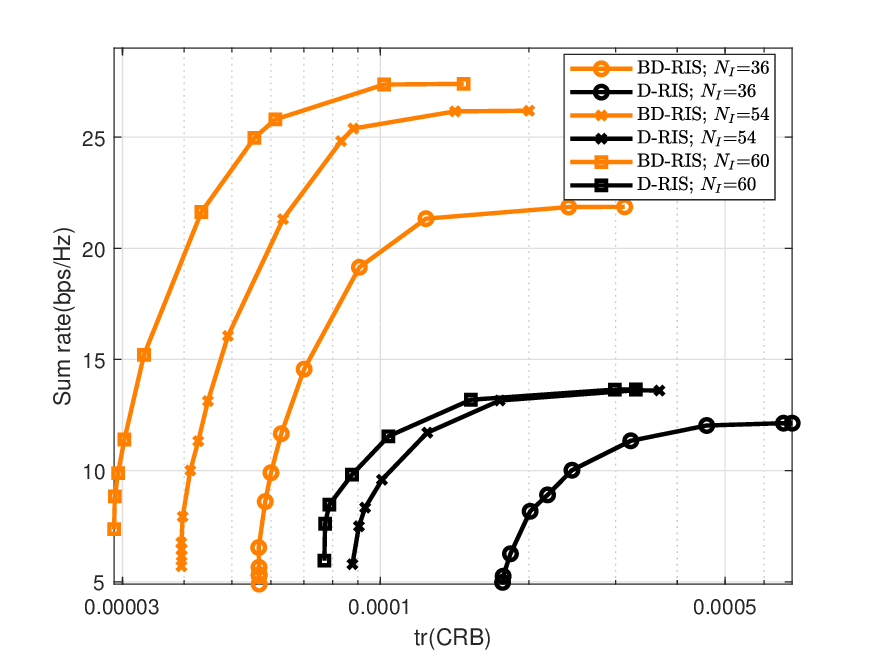}
        \vspace{-3mm}
		\caption{The trade-offs under different numbers \protect\\ of BD-RIS elements. $K=4$, $Q=1$, $N_{T}=6$, \protect\\$N_{S}=8$, $M=128$, $P=20\ \mathrm{dBm}$.}
		\label{RIS_element} 
	\end{minipage}
	\begin{minipage}[t]{0.34\linewidth}
		\centering
		\includegraphics[width=2.4in, height=1.95in]{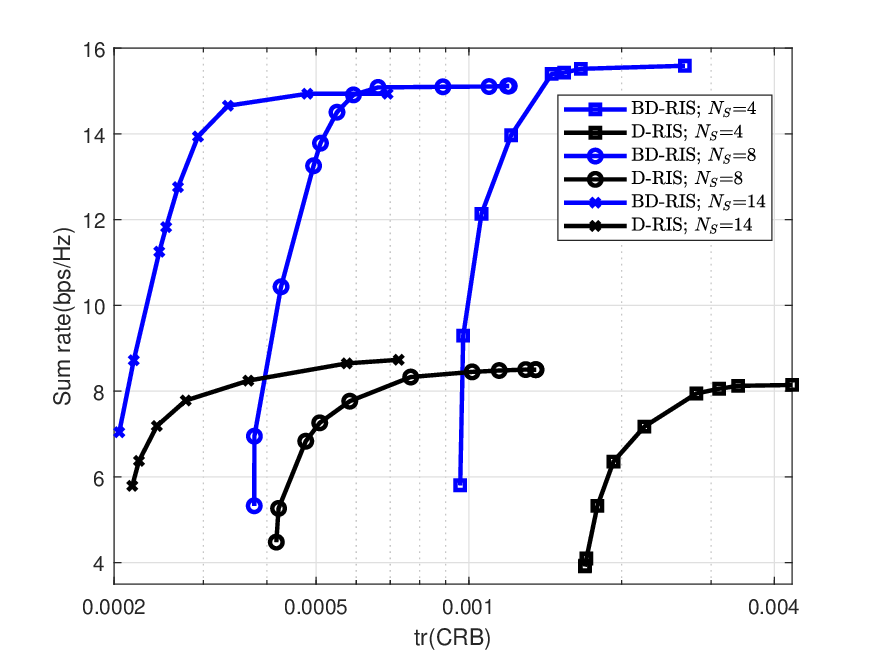}
        \vspace{-7.3mm}
        \caption{The trade-offs under different
			numbers \protect\\of sensor elements. $K=4$, $Q=1$, $N_{I}=32$, \protect\\$N_{T}=4$, $M=64$, $P=15\ \mathrm{dBm}$.} 
		\label{Sensor_element}
	\end{minipage}%
	\begin{minipage}[t]{0.34\linewidth}
		\centering
		\includegraphics[width=2.2in, height=1.95in]{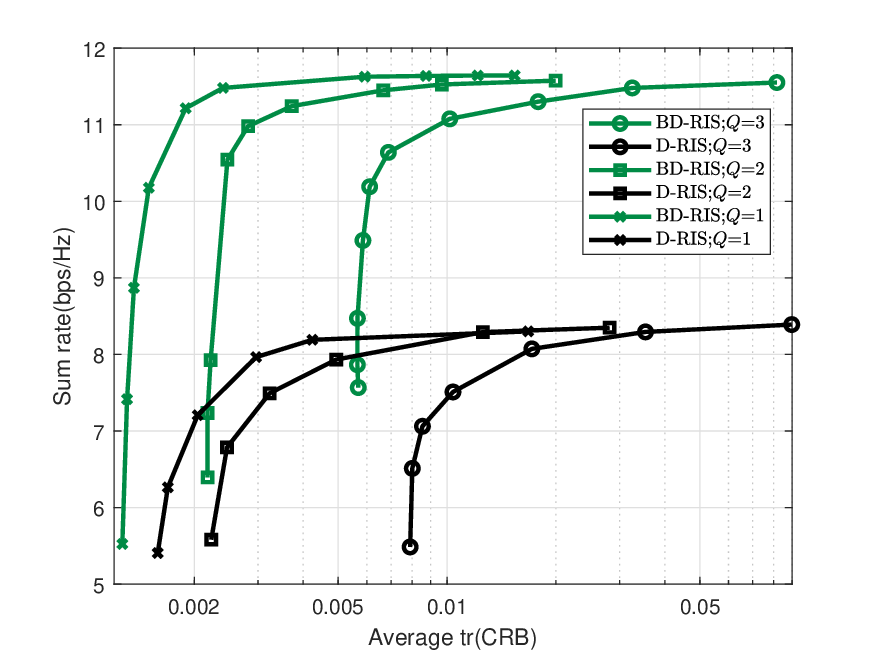}
        \vspace{-3mm}
		\caption{The trade-offs under different
			 numbers \protect\\of targets. $K=2$, $N_{I}=16$, $N_{T}=4$, $N_{S}=4$,\protect\\ $M=64$, $P=20\ \mathrm{dBm}$.}
		\label{Target_number}
	\end{minipage}
%\vspace{-3mm}
\end{figure*}
\section{Numerical Results}
In this section, we evaluate the performance of the proposed transmitter-side BD-RIS for mmWave ISAC. The sum rate $\sum_{k=1}^K R_k$ and the trace of the CRB are selected for evaluating communication and sensing performance, respectively.
% the CRB metric
Note that when there are multiple sensing targets, the average tr(CRB) defined by $\mathrm{tr}(\mathbf{F}^{-1})/Q$ is chosen to ensure comparison fairness.
% the location
Unless otherwise specified, we consider $K=4$ communication users and $Q=1$ sensing target. We assume that the BD-RIS is 10 wavelengths away from the active antennas, with the carrier frequency set to $f_c = 30$ GHz.  
The communication users are randomly placed at $d_k = 30 \sim 50$ m away from the BD-RIS, where the communication channels $\mathbf{h}_k$ are modeled as the Rician channels with the same parameter setting as in \cite{wzl23_STARS}.
And we consider $3$ different targets named target 1-3.  They are located at $[d_q, \theta_q, \phi_q]=[30\ \mathrm{m},30^{\circ} ,30^{\circ} ]$, $[45\ \mathrm{m},15^{\circ} ,60^{\circ} ]$, and $[40\ \mathrm{m},60^{\circ} ,-45^{\circ}]$ with respect to the BD-RIS, respectively. The signal attenuation variance for sensing is calculated as $\sigma^2_{\alpha_q}=\frac{\beta \lambda^2}{(4\pi)^3 d_q^4}$, where $\beta$ denotes the radar cross section (RCS) and it is set to 5. 
The total power budget at the active antennas is $P=20\ \mathrm{dBm}$, and the noise power is given as $\sigma_c^2=-60\ \mathrm{dBm}$ and $\sigma_r^2=0\ \mathrm{dBm}$ \cite{xj23_sigma}. Each CPI contains $N=128$ transmission blocks. 

To verify the efficiency of the transmitter-side BD-RIS-aided ISAC system, we select the conventional D-RIS-aided ISAC system as the baseline, where the D-RIS operates in the transmissive mode  as well \cite{dwn23_ITScommunication}. Let $\theta_{T,i}, i=[1, \dots, N_I]$ denote the phase shift angle of the conventional D-RIS. The corresponding diagonal scattering matrix $\bm{\Psi}=[e^{j\theta_{T,1}}, \dots, e^{j\theta_{T,N_I}}]$ can then be obtained via the same method as stated in \cite{fty24_BDRISlowcomplexity}.
%via $\bm{\Psi}=\mathrm{symuni}(\mathbf{I}_{N_I}\odot \bm{\Psi}^\ast)$ \cite{fty24_BDRISlowcomplexity}, 
%where $\odot$ denotes the Hadamard product. $\bm{\Psi}^\ast$ is the optimal solution on the relaxed set $\mathcal{M}$ mentioned in (\ref{psi_relaxed}). 
The numerical results are obtained by averaging over 100 realizations.

In Fig. \ref{RIS_element}, we consider $N_T=6$ active antennas and $N_S=8$ sensor elements. We study the communication sum rate versus the sensing tr(CRB) for different numbers of BD-RIS elements $N_I$. 
As expected, better communication performance, i.e., the higher sum rate, as well as better sensing performance, i.e., the lower tr(CRB), are obtained with increasing BD-RIS elements. 
BD-RIS consistently achieves a better communication and sensing trade-off performance than the conventional D-RIS. 
Moreover, the trade-off gain of BD-RIS over D-RIS is more explicit with increasing number of BD-RIS elements. This is attributed to the growing flexibility in BD-RIS design as the number of RIS elements increases.

In Fig. \ref{Sensor_element}, we illustrate the influence of the number of the sensor elements $N_S$, where $N_I=32$, $N_T=4$, $M=64$ and $P=15\ \mathrm{dBm}$. 
It can be observed that as the number of sensor elements increases, the sensing performance becomes better, i.e., tr(CRB) becomes lower. Owing to the additional degree of freedom (DoF) introduced by the fully-connected passive elements to construct the directional sensing beam, BD-RIS dramatically enhances the sensing performance. 
%With more emphasis on the sensing functionality, the communication sum rate for the BD-RIS meets unpromising decrease owing to its complicated beyond-diagonal scattering matrix design. But its communication performance still shows superiority than conventional ITS.

The trade-offs under different numbers of targets are demonstrated in Fig. \ref{Target_number}, where $Q=1$ (target 1), $Q=2$ (target 1-2) and $Q=3$ (target 1-3) are considered. There exists $K=2$ communication users with $N_I=16$, $N_T=4$, $N_S=4$ and $M=64$. As observed, BD-RIS shows an explicit trade-off gain over the conventional D-RIS under different target numbers. The communication and sensing performance of both BD-RIS and the conventional D-RIS deteriorates as the number of targets increases. This is because less beamforming power is allocated to each sensing target. Surprisingly, BD-RIS shows the capability to detect more sensing targets than the conventional D-RIS while maintaining the same sum rate performance for communication users. This further shows the great potential of integrating BD-RIS into mmWave ISAC systems.

\section{Conclusion}
In this work, we propose a novel BD-RIS-aided transmitter architecture to reduce the power consumption and improve the communication and sensing performance for mmWave ISAC. 
The digital beamforming matrix of the transmitter and the scattering matrix of the BD-RIS are designed using the proposed two-stage algorithm with the aim of jointly maximizing the communication sum rate and minimizing the largest eigenvalue of the CRB matrix of the sensing targets.
% the simulation
Numerical results demonstrate that the proposed transmitter-side BD-RIS significantly enhances the trade-off gain between communication and sensing compared to the conventional D-RIS. The results show the great potential of integrating BD-RIS into mmWave ISAC systems.

\section*{Appendix A} \label{appendix A}
We define $\mathbf{u}_{r}[m]=\mathbf{y}_{r}[m]-\mathbf{z}_{r}[m]=\mathbf{B}\mathbf{U}\mathbf{A}^{T}\bm{\Psi}\mathbf{H}\mathbf{x}[m]$. With its partial derivative of $\mathrm{Re(\alpha_i)}$, i.e., $\mathbf{B}\mathbf{e}_{i}\mathbf{e}_{i}^{T}\mathbf{A}^{T}\bm{\Psi}\mathbf{H}\mathbf{x}[m]$, $\forall i \in \mathcal{Q}$, the element of $\mathbf{F}_{\mathrm{Re}(\bm{\alpha})\mathrm{Re}(\bm{\alpha})}\in \mathbb{R}^{Q\times Q}$ in (\ref{fisher}) is calculated as (see, e.g., [13, Appendix A])
\begin{equation}
    \begin{aligned}
        &\mathbf{F}_{\mathrm{Re}(\alpha_i)\mathrm{Re}(\alpha_j)}=\frac{2}{\sigma_r^2}\mathrm{Re}\bigg[\mathrm{tr}\bigg\{ \sum_{m\in\mathcal{M}} \frac{\partial\mathbf{u}_{r}^H[m] }{\partial \mathrm{Re}(\alpha_i)}\frac{\partial\mathbf{u}_{r}[m]}{\partial \mathrm{Re}(\alpha_j)}\bigg \}\bigg]\\
	=&\frac{2M}{\sigma_r^2}\mathrm{Re}\Big \{\mathbf{e}_{j}^{T}\mathbf{A}^T\bm{\Psi}\mathbf{H}\mathbf{R}_X\mathbf{H}^H\bm{\Psi}^H\mathbf{A}^\ast   \mathbf{e}_{i} \mathbf{e}_{i}^{T} \mathbf{B}^H\mathbf{B}    \mathbf{e}_{j} \Big \} \\
	=&\frac{2M}{\sigma_r^2}\mathrm{Re}\Big \{(\mathbf{A}^{H}\bm{\Psi}^\ast\mathbf{H}^\ast\mathbf{R}_{x}^{\ast }\mathbf{H}^T\mathbf{\Psi}^T\mathbf{A})_{ij}(\mathbf{B}^{H}\mathbf{B})_{ij}  \Big \}, \forall i,j\in \mathcal{Q},
    \end{aligned}
\end{equation}
where $(\cdot)_{ij}$ denotes the element in the $i$th row and $j$th column. Other terms of FIM can be calculated in the same way, which are omitted here due to space limitation.

\section*{Appendix B} \label{appendix B}
With $\mathbf{X}=\mathbf{V}^H_1\bm{\Psi}\mathbf{U}_2 \in \mathbb{C}^{N_I \times N_I}$, $\mathbf{\Sigma}_1=\mathbf{S}_1^H\mathbf{S}_1 \in \mathbb{C}^{N_I \times N_I}$ and $\mathbf{\Sigma}_2=\mathbf{S}_2\mathbf{S}_2^H \in \mathbb{C}^{N_I \times N_I}$ defined in Section \ref{psi section}, we have
\begin{equation}
    \begin{aligned}
&\mathrm{Re}\left\{\mathrm{tr}\left(\mathbf{\Sigma}_1\mathbf{X}\mathbf{\Sigma}_2\mathbf{X}^H\right)\right\}=\sum_{i=1}^{N_I}\mathrm{Re}\left\{\mathrm{tr}\left[ \mathbf{\Sigma}_1(\mathbf{\Sigma}_2)_{ii}\mathbf{x}_i\mathbf{x}_i^H \right] \right\}\\
=&\sum_{i=1}^{N_I}\sum_{j=1}^{N_I}(\mathbf{\Sigma}_1)_{jj}(\mathbf{\Sigma}_2)_{ii}|x_{ji}|^2 \leq \sum_{i=j=1}^{N_I}(\mathbf{\Sigma}_1)_{jj}(\mathbf{\Sigma}_2)_{ii},
    \end{aligned}
\end{equation}
where $\mathbf{x}_i=[x_{1i}, \dots, x_{N_Ii}]^T \in \mathbb{C}^{N_I \times 1}$ denotes the $i$th column of the matrix $\mathbf{X}$.
The inequality holds since $(\mathbf{\Sigma}_n)_{ii}\geq (\mathbf{\Sigma}_n)_{jj}\geq 0, i\leq j, n\in\{1,2\}$, and $\mathbf{X}\in \mathcal{M}$. It is with equality if and only if $\mathbf{X}=\mathbf{I}_{N_I}$.
\bibliographystyle{IEEEtran}  
\bibliography{ref}

\end{document}